\documentclass[prb,twocolumn]{revtex4-1} 

\usepackage{amsmath}  
\usepackage{amsfonts} 
\usepackage{graphicx} 

\begin{document}


\title{On Milne--Barbier--Uns\"old relationships}

\author{Fr\'ed\'eric Paletou} \email{frederic.paletou@univ-tlse3.fr}
\affiliation{Universit\'e Paul Sabatier, Observatoire
  Midi--Pyr\'en\'ees, Cnrs, Cnes, Irap, F--31400 Toulouse, France}



\date{\today}

\begin{abstract}
This short review aims to clarify upon the origins of so--called
Eddington--Barbier relationships, which relate the emergent specific
intensity and the flux to the photospheric source function at specific
optical depths. Here we discuss the assumptions behind the original
derivation of Barbier (1943). We also point to the fact that Milne had
already formulated these two relations in 1921.
  \end{abstract}

\maketitle 

\section{Introduction}

The theory of radiation transfer is fundamental in
astrophysics. Besides the \emph{in situ} exploration of various bodies
in the solar system, cosmic rays astrophysics and the recent
spectacular advent of gravitational wave detections, remote sensing of
radiation remains the primary means through which we advance our
knowledge of celestial bodies.

Analytical solutions were successively derived, mostly during the
first half of the XX\emph{th} century, and beyond the advent of
numerical computing, into the 1960's. After pioneering contributions
of Schuster (1905) and Schwarzschild (1906), important results were
further established, in particular for the case of stellar
photospheres in radiative equilibrium. They relate, for instance, to
the temperature distribution in a so-called ``gray atmosphere'', and
to the associated limb--darkening law of radiation.  One may also
mention the $\sqrt{\varepsilon}$ surface value for simplified
radiation transfer \emph{out of} local thermodynamical equilibrium
(see e.g., Hubeny 1987; Lambert et al. 2016).

Fundamental elements of radiative transfer can be found in the texts
of Rutten (2003) and Hubeny \& Mihalas (2014).

\section{Eddington--Barbier relationships}

The so-called Eddington--Barbier relationships constitute fundamental
analytic results, systematically presented in most textbooks and
lectures about radiative transfer in astrophysics. In most cases,
they are introduced and derived assuming that the source function is
just a \emph{linear} function of the optical depth:

  \begin{equation} 
S(\tau)= a + b \tau  \, ,
  \end{equation} 
where $a$ and $b$ are arbitrary coefficients.

It is then easy to derive the \emph{emergent} specific intensity, from
a plane-parallel semi-infinite atmosphere, according to:
  
\begin{equation}
 I(\mu)= 
 \int_{0}^{\infty}   { S(\tau) e^{-(\tau / \mu)}
 \left( d\tau/\mu \right) }  \, .
  \label{eq:emergent}
\end{equation}
Here $\mu$ is the usual cosine of the angle of the ray to the vertical
direction. Given equation (1), the specific intensity is merely:

  \begin{equation} 
I(\mu)= a + b \mu  \, ,
  \end{equation} 
that is, the source function at optical depth $\tau=\mu$ i.e.,

  \begin{equation} 
I(\mu)= S(\tau=\mu)  \, .
  \end{equation} 

In other words, this means also that, for a given line of sight the
emergent intensity equals the source function at a depth found after
crossing an optical depth unity \emph{along the line of sight}.
  
A second relationship can also be derived for the \emph{emergent flux}
defined as:

\begin{equation}
 {\cal{F}} = 2 \pi
 \int_{0}^{1} { I(\mu) \mu d\mu } \, .
  \label{eq:flux}
\end{equation}
The quantity ${\cal{F}}$ is relevant to spatially \emph{unresolved}
objects, like most stars (besides the Sun), and it is easy to show
that the emergent flux is then characterized by the source function at
optical depth $\tau=2/3$.

Many commonly-read textbooks such as Athay (1972), Mihalas (1978), the
very popular $e-$book of Rutten (2003), and even the recent Hubeny \&
Mihalas (2014) omit however to cite any original publication
establishing first these two classical relationships.
    
\section{Original derivation}

The origin of the derivation of these relationships can be found in an
article of French astronomer \emph{Daniel Barbier} published in 1943,
although he did not address explicitly the case of the specific
intensity there. The original derivation of Barbier starts with the
following Taylor series expansion for the source function:

\begin{equation}
  S(\tau)=S(\tau_{*}) + (\tau-\tau_{*})S^{\prime} (\tau_{*})
  + {1 \over 2} (\tau-\tau_{*})^2 S^{\prime\prime} (\tau_{*})
  \label{eq:taylor}
\end{equation}
that we truncate here at 2nd order. In this expression $S^{\prime}$
and $S^{\prime\prime}$ are respectively the first and the second
derivatives vs. $\tau$ of the source function. This expansion is
introduced into Eq.\,(\ref{eq:emergent}), and straightfoward
integrations give the following expression for the emergent specific
intensity:

\begin{eqnarray}
  I(\mu) & = & S(\tau_{*}) + (\mu-\tau_{*})S^{\prime} (\tau_{*}) \nonumber \\ 
  & + & (\mu^2 - \mu \tau_{*} + \frac{1}{2} \tau_{*}^2) S^{\prime\prime}
  (\tau_{*}) \, .
  \label{eq:Barbier}  
\end{eqnarray}

In his original article of 1943, D. Barbier does \emph{not} give an
expression for the emergent specific intensity, but does give the
emergent ``total flux'', where intensity is integrated over $\mu$
(equation 5). However, we can adopt his argument, and \emph{choose
  $\tau_{*}$ which makes the term in $S^{\prime}$ vanish, and which
  minimizes that in $S^{\prime\prime}$}.  It is therefore obvious
that:

\begin{equation}
  \tau_{*}=\mu \, .
\end{equation}
In such a case, the emergent specific intensity is:

\begin{equation}
  I(\mu)=S(\tau=\mu) +
   \frac{1}{2} \mu^2 S^{\prime\prime} (\tau=\mu) \, ,
  \label{eq:BarbierSolI}
\end{equation}
which is indeed identical to $S(\tau=\mu)$, if one assumes that the
source function is no more than \emph{linear} in the optical depth, so
that $S^{\prime\prime}(\tau)=0$.

\section{Discussion}

Barbier (1943) cites Eddington quite precisely, pointing at
\emph{p. 330} of his famous textbook \emph{The internal constitution
  of stars} (1926). In this chapter, \emph{The Outside of a Star},
Eddington states that the \emph{effective temperature} of the
angle-dependent atmospheric radiation should be the temperature of the
layer where $\tau \approx \mu$. This may have inspired Barbier to
adopt a Taylor series expansion method.

Soon after Barbier's contribution, Uns\"old (1948; 1949, in English)
makes explicit these classical relationships, \emph{both} for the
specific intensity and for the emergent flux. Uns\"old (1955) gives
direct credit to Barbier in his famous textbook, \emph{Physik der
  Sternatmosph\"aren}. However, he wrongly cites Barbier (1944),
instead of Barbier (1943)! About Barbier's method based on a Taylor
series expansion of the source function $S$ around a certain optical
depth $\tau_{*}$ which has to be determined, he writes, originally in
German, that: \emph{``a method of approximation, proposed by
  A.S. Eddington and better argumented by D. Barbier is still very
  useful and interesting''}. Uns\"old coins it also the
\emph{``$x=\cos \vartheta$--Methode''} of Eddington and Barbier, where
$x$ is used for optical thickness $\tau$, and $\mu=\cos \vartheta$.

Kourganoff's (1952) textbook gives a proper citation and description
of Barbier's original contribution in his \S18.2: \emph{``After giving
  Barbier's demonstration, which is known as the $\tau_{*}$-method''},
which is however immediately followed by: \emph{``we shall explain why
  it seems to us to be unsatisfactory''...} Modern texts reflect
Kourganoff's clear statement that: \emph{``Now Barbier's demonstration
  (or a direct calculation) shows that all of the Eddington--Barbier
  relations are rigorously true if the source function is a linear
  function of $\tau$''.} And despite a critical discussion in the
remaining of \S18, Kourganoff concludes with: \emph{``The
  Eddington--Barbier relations, apart from the applications which have
  already been made by Barbier himself and by Uns\"old, are extremely
  useful when one wants to represent the connexion between the source
  function and certain observable quantities like $I(0,\mu)$ and
  ${\cal{F}}(0)$''.}

With time, citations to Barbier (1943) vanish, although they appear in
textbooks such as those of Zirin (\S6.10., 1988) and Castor (\S5.7.,
2004). This is, surprinsingly, \emph{not} the case for the famous and
comprehensive Mihalas (1978) textbook. However, in its \S2-2 an
exercise (2-5) is directly inspired by the method used by Barbier
(1943). It may be unfortunate though, that the revised and expanded
textbook by Hubeny \& Mihalas (2014) does not elaborate on the
original derivation of one of the most famous result of analytical
radiation transfer.

\section{A lost contribution of Milne?}

During the course of our investigations on the original contributions
leading to the so-called Eddington--Barbier relations, we also went
back to an article of V.V. Ivanov (1991) in the proceedings of the
Trieste conference \emph{Stellar Atmospheres: beyond classical
  models}.

At the end of section \emph{History of ART}, where ART stands for
``Analytical Radiative Transfer'', Ivanov writes this somewhat
intriguing statement: \emph{``The standard Eddington approximate form
  of the temperature distribution in a grey atmosphere,}

\begin{equation}
  T^4 = (3/4) T_{\rm eff}^4 (\tau + 2/3) \, ,
  \label{eq:greyatmos}
\end{equation}
\emph{belongs not to Eddington, but to Milne. In 1917, he introduced
  the approximation known today as the Eddington approximation.''}

However, it seems that Milne did not publish any astrophysical result
before 1921, according to Tayler (1996) for instance. This led us to
look back with some care to the early contributions of Milne in the
domain of radiative transfer and stellar atmospheres. And his first
article of 1921, \emph{Radiative equilibrium in the outer layers of a
  star: the Temperature Distribution and the Law of Darkening}
contains in Eqs. (36) and (37) both Eddington--Barbier relations, for
the specific intensity and for the flux, as shown in Fig. (1).

\begin{figure}[]
  \includegraphics[width=8.25 cm,angle=0]{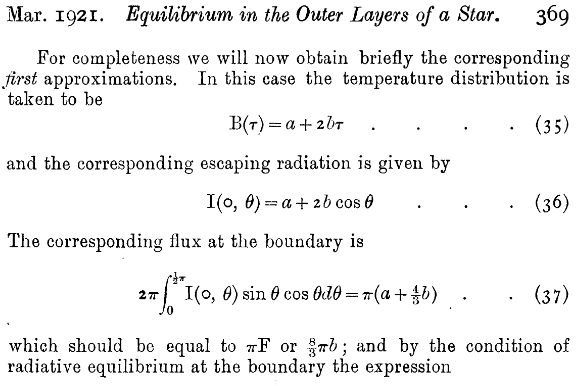}
  \caption{In this excerpt of Milne's (1921) article, $B(\tau)$ should
    also read as the source function and Eqs. (36) and (37) are just
    the Eddington--Barbier relationships respectively on the specific
    intensity and on the flux, published more than twenty years before
    Barbier (1943).}
  \label{Fig1}
\end{figure}

In this article, the derivation is formally distinct from the one used
by Barbier (1943). Milne is inspired by the method already used by
Schwarzschild and Jeans. He uses, in particular, simplifications for
lower and upper boundaries incident radiation which are assumed to be
distinct but independent of direction. Then he gives an \emph{integral
  form} of the ``temperature distribution'' for a semi--infinite
atmosphere, with no incoming radiation on its surface. First, he shows
that a linear source function is a possible solution at \emph{large}
optical depths, and even at this stage the first part of his Eq.\,(24)
already contains the Eddington--Barbier relationship for the specific
intensity, when (inappropriately though) setting... $\tau=0$. He goes
however further and implements a method of successive approximations,
which leads to a second approximation for the source function of the
form:

\begin{eqnarray}
  S(\tau) & = & a+2b\tau + \frac{1}{2}e^{-\tau}(b-a-b\tau) \nonumber \\ 
  & + & \frac{1}{2}(a\tau+b\tau^2) \int_{\tau}^{\infty} {{{e^{-y}} \over {y} }  dy}
  \, .
  \label{eq:Milne}  
\end{eqnarray}
This expression has indeed the same form as the first approximation,
linear in $\tau$, for large optical depths but now, the first
exponential integral $E_1(\tau)$ appears, as in the
Schwarzschild--Milne equations (see e.g., Hubeny \& Mihalas 2014). It
is also more accurate in the sense that it exhibits the often
forgotten \emph{singularity} in $dS/d\tau$ at $\tau=0$ (see e.g.,
Chevallier et al. 2003 and references therein).

Both the so-called Eddington--Barbier relationships appear once Milne,
\emph{``for completeness''}, goes back to his first approximation,
when the source function is just linear in the optical depth. At this
stage, Milne does \emph{not} comment any further on his Eqs. (36) and
(37). This may also explain why these intermediate results of Milne
were neither used, nor cited further. Perhaps they may also have been
eclipsed by other aspects of Milne's important contributions in the
early 1920's?

For instance, following the remark of Ivanov (1991), in his first 1921
article Milne establishes also a limb--darkening law such that:

\begin{equation}
  \frac{I(0,\mu)}{I(0,1)} = \frac{3}{5} (\mu + 2/3) \, ,
  \label{eq:limdrk}
\end{equation}
improving on Schwarzschild and Jeans. It is also fully consistent with
the assumption known as the ``Eddington approximation'', which leads
to the (constant) $2/3$ appearing in Eq.\,(\ref{eq:greyatmos}).

\section{Conclusion}

One may argue that Uns\"old (1948; 1949, in English) makes the
classical relations explicit, for the first time, \emph{both} for the
specific intensity and for the emergent flux, following Barbier
(1943). However they were \emph{both} expressed in Milne (1921), but
cited neither by Barbier, nor by Uns\"old (or anyone else, to the best
of our knowledge) more than twenty years after.

After our investigations, we would therefore propose to the
astrophysical community to shift, at last, from the
``Eddington--Barbier'' usual designation to the \emph{fairer}
``Milne--Barbier--Uns\"old'' relationships.

Finally, we also report that despite its legacy, Barbier (1943) has
been cited only four times so far, according to the ADS service! This
is also the case for Uns\"old (1948; and it is even worse for his 1949
article, although written in English)...

\begin{acknowledgments}
This review was initiated after a question of one of
\emph{Universit\'e de Toulouse} astrophysics master student, Mrs
Alexandra Le Reste, during a lecture of Sept. 2017. We also wishes to
thank Prof. R.J. ``Rob'' Rutten for fruitful and enjoyable
discussions, as well as Drs C\'eline Reyl\'e (\emph{Observatoire de
  Besan\c con}, France) and Torsten B\"ohm (CNRS, IRAP, Toulouse,
France) for their kind assistance, and finally Dr. Ivan Hubeny, for
his encouragement to make available this note in English. This
research has made use of NASA's Astrophysics Data System, a fantastic
tool to search and read again the historical astrophysical literature.
\end{acknowledgments}

\end{document}